\documentclass{ws-ijmpd}
\usepackage[super,compress]{cite}
\begin{document}

\markboth{Ghosh, Bhadra and Mulhopadhyay}
{ (Gravitational lensing study of cold dark matter led galactic halo)}

\catchline{}{}{}{}{}
%
%%%%%%%%%%%%%%%%%%%%%%%%%%%%%%%%%%%%%%%%%%%%%%%%%%%%%%%%%%%%%%%%%%%%

\title{Gravitational lensing study of cold dark matter led galactic halo %\footnote{For the title, try not to use more than 3 lines.
%Typeset the title in 10~pt Times roman, uppercase and boldface.}  
}

\author{Samrat Ghosh
%\footnote{Typeset names in 8~pt roman, uppercase. Use the footnote to indicate the present or permanent address of the author.}
}

\address{High Energy $\&$ Cosmic Ray Research Centre, University of North Bengal, \\
Siliguri, West Bengal, 734013,
India \\
%\footnote{State completely without abbreviations, the affiliation and mailing address, including country. Typeset in 8~pt Times italic.}\\
samrat.ghosh003@gmail.com}

\author{Arunava Bhadra}

\address{High Energy $\&$ Cosmic Ray Research Centre, University of North Bengal, \\
Siliguri, West Bengal, 734013,
India \\
aru\_bhadra@yahoo.com}

\author{Amitabha Mukhopadhyay}

\address{Department of Physics, University of North Bengal,\\
Siliguri, West Bengal, 734013,
India \\
amitabha\_62@rediffmail.com}

\maketitle

\begin{history}
\received{Day Month Year}
\revised{Day Month Year}
\end{history}

\begin{abstract}
In this work the space-time geometry of the halo region in spiral galaxies is obtained considering the observed flat galactic rotation curve feature, invoking the Tully-Fisher relation and assuming the presence of cold dark matter in the galaxy. The gravitational lensing analysis is performed treating the so obtained space-time as a gravitational lens. It is found that the aforementioned space-time as the gravitational lens can consistently explain the galaxy-galaxy weak gravitational lensing observations and the lensing observations of the well-known Abell 370 and Abell 2390 galaxy clusters.
\end{abstract}


\ccode{PACS numbers:}

%\tableofcontents

\section{Introduction}
%\label{intro}
The astrophysical observations reveal that after the termination of the luminous disk the expected Keplerian fall-off is absent in rotation curves (variation of the angular velocity of test particles with distance from the galactic center) of spiral galaxies  \cite{rob73, rub78, rub79, per96, sof01}. The frequency shift of the 21 cm HI emission line from neutral hydrogen cloud at large distances from the galactic center rotating in circular orbits allows constructing a rotation curve of galaxies involving distances up to a few tens of kpc or even a few hundreds of kpc in few cases. The observed flatness of galactic rotation curves implies that either the galaxies contain far more matter than contributed by the luminous matters such as stars, planets, and gas or the law of gravity is different at large distances. The existence of invisible matter (commonly referred to as dark matter) is also supported from the observed velocity dispersion of galaxies in the galactic clusters \cite{zwi33}, \cite{luk14}, and gravitational lensing by galaxies \cite{luk14, bel92, mao93, kin98, mar04, wae00, wit00}. 
 
The dark matter hypothesis has also received support from cosmological observations. The $\Lambda$CDM model, where $\Lambda$ is the cosmological constant and CDM stands for cold dark matter, fits the cosmological observations well and is quite successful in describing the formation and evolution of the large scale structure in the Universe (see for instance \cite{ade16, pat16}). In cosmology, the cold dark matter hypothesis draws from two phenomena - inflation and big-bang nucleosynthesis. The inflationary idea suggests that the Universe is nearly flat with matter density equals to critical density which receives support from the observed anisotropy features in the cosmic microwave background radiation (CMBR) \cite{spe03}. The baryon density inferred from nucleosynthesis suggests that ordinary matter can contribute at most $15\%$ of the critical density \cite{oli00}. Hence, if the inflationary picture is correct, then most of the matter in the Universe must be nonbaryonic. The $\Lambda$CDM model interprets that the gravitational attraction of cold dark matters leads to the formation of cosmic structures, and it also plays important role in holding the structures together. 
  
The space-time geometry of galactic halo in presence of dark matter is a very relevant issue. Besides the study of the effects of gravitational interactions in the galactic halo region it also offers the possibility of cross-verification of the existence of dark matter itself through different local gravitational phenomena such as gravitational lensing, gravitational time delay \cite{sar16}, time advancement \cite{bha10, gho15, gho19} etc. A naive Newtonian analysis suggests that the tangential velocity ($v_{\varphi}$) of rotation $\beta_{\varphi}=\sqrt{\frac{GM}{c^2 r}}$, where $\beta_{\varphi} = v_{\varphi}/c$, c is the speed of light, $G$ is the gravitational constant, $M$ is the total mass inside the radius $r$ of the galaxy and $r$ is the distance from the galactic center. The observed flatness of galactic rotation curves implies that $M$ is a function of $r$ that increases linearly with $r$.

Several attempts were made to model dark matter halos in the general relativistic framework. In the Newtonian approach gravitational field is solely represented by gravitational potential, the matter density solely plays the role of generating the gravitational potential which can be completely determined by the observed rotation curve in the galactic halo region. In contrast, even in the spherically symmetric situation, general-relativistic analysis requires knowledge of two metric coefficients ($g_{tt}$ and $g_{rr}$), to completely describe the gravitational field of the galactic halo. One of the underlying reasons for such a difference is that in the general relativistic framework pressure also contributes to gravitational field unlike in the Newtonian approach. While the $g_{tt}$ can be obtained from the features of the rotation curve, additional input about the equation of state of dark matter is required to determine $g_{rr}$. 

In the present work, we shall obtain a general relativistic solution of space-time geometry of galactic halo in presence of CDM assuming the flatness of the rotation curve as observed. We shall also invoke the Tully-Fisher relation to fix the velocity parameter as a function of luminous mass. Note that our objective is not to model the dark matter of the galaxy, rather we shall derive the space-time metric in the galactic halo region taking the observed feature of the galactic rotation curve as input and assuming the presence of cold dark matter in the galaxy.  The gravitational lensing observations provide compelling evidence for the existence of dark matter in the galactic halo. We shall study the gravitational lensing due to the space-time metric of the galactic halo as derived in the present work and compare it with the observations to examine the consistency of the model.   

The organization of the paper is as the following. In the next section, we shall evaluate the gravitational potential at galactic halo exploiting the observed flat rotation curve feature and considering the presence of cold dark matter. We shall discuss other relevant issues like the Tully-Fisher relation, stability of circular geodesics in section 3. In section 4 we shall study gravitational lensing due to the derived space-time metric. We shall discuss our results in section 5 and finally conclude in the same section.

\section{The space time geometry of galactic halo} 
%\label{sec:1}
In Newtonian gravity, the tangential velocity of a test particle in circular orbits around the central mass distribution is obtained simply by equating the centripetal acceleration with the gravitational acceleration due to central mass that leads to $v_{\varphi}=\sqrt{\frac{GM}{r}}$,  The determination of tangential velocity in GR framework is slightly complex. Assuming that the galactic halo is spherically symmetric, the general static space-time metric of the halo can be written in curvature coordinates as \cite{mis73}

\begin{equation}
ds^{2}=-e^{2\lambda(r)}dt^{2}+ \frac{dr^2}{1-\frac{2m(r)}{r}}+ r^{2}(d\theta^2+sin^2\theta d\phi^2),
\end{equation}

where $\lambda(r)$ and $m(r)$ are functions of $r$ only. We are expressing quantities in natural units i.e. $c$ and $G$ are taken as 1. The function $\lambda(r)$ is known as the “potential” and $m(r)$ is the “shape function” which essentially reflects the effective gravitational mass. 

\subsection{Galactic potential from the flat rotation curve feature} 

Assuming that test particles move on the equatorial plane ($\theta=\pi/2$) the tangential velocity of a non-relativistic test particle in a circular orbit can be obtained from the study of geodesics for the above space-time metric which is given by \cite{mat00, rah10} 

\begin{equation}
\beta_{\varphi}^{2}=r \lambda^{\prime}(r)
\end{equation}

where prime denotes the derivative with respect to r. Since observations suggests that $\beta_{\varphi}$ is nearly constant at large galactic distances, the above equation immediately gives at halo region $e^{2\lambda(r)} \propto r^{2 \beta^{2}_{\varphi}}$. This form of $g_{tt}$ is adopted in the several previous works \cite{mat00, mat00a, nuc01, nuc00, fay04, rah10, arb01, rah14} for galactic halo.

%\begin{equation}
%e^{\lambda(r)} \propto r^{2 \beta^{2}_{\varphi}} \;.
%\end{equation}

Here we look for a form of $g_{tt}$ that can be recast as perturbation of Minkowski metric ($g_{\mu\nu} = \eta_{\mu\nu} + h_{\mu\nu}$, where $ h_{\mu\nu}$ is the small perturbation over the Minkowski metric $\eta_{\mu\nu}$ \cite{car19}) as expected in the weak gravitational field regime of the galactic halo region. Accordingly we write $g_{tt}$ of Eq.(1) as $e^{2\lambda(r)} = 1 -\frac{2M_B}{r} + f(r) $, where $f(r) (\equiv h_{tt})$ is a function of r that arises due to the presence of dark matter and $M_B$ is the total baryonic matter of the galaxy within radius r. The observations suggest that after the galactic bulge the density of baryonic matter is very small and $M_B$ may be taken as a constant. When r is small, $f(r)$ is smaller than $\frac{2M_B}{r}$ and vice versa for large r. Therefore, in the galactic halo region we can approximate $e^{2\lambda(r)} \simeq 1 + f(r)$. Since the magnitude of $\beta^{2}_{\varphi}$ is very small we shall keep only the leading order terms in $\beta^{2}_{\varphi}$ in the solution of $\lambda(r)$ ignoring higher order terms in $\beta_{\varphi}$. When flat rotation curve feature is invoked, we get the following solution 
\begin{equation}
f(r) \simeq  2 \beta^{2}_{\varphi} ln r + C1
\end{equation}

where $C1$ is an integration constant which may be fixed from the boundary conditions. 
Here it is worthwhile to mention that the modified Newtonian dynamics (MOND) model \cite{mil83}, which is probably the most viable alternative to the dark matter hypothesis, also essentially employs logarithmic potential \cite{mil13}. At this stage i.e. just from the flat rotation curve feature one cannot rule out the possibility that some modification of general relativity could be the origin of the logarithmic form in the potential.  

\subsection{The cold dark matter effect on space-time}

For a complete understanding of space-time geometry in the halo region, knowledge about $g_{rr}$ is also required. Additional input in the form of dark matter equation of state is needed to determine $g_ {rr}$. The nature of dark matter is an unanswered issue of contemporary astrophysics. The only information available about dark matter is that it has not shown any interaction with the baryonic matter except the gravitational interaction. 

Numerical simulations of structure growth suggest that the dominant part of the dark matter in the universe is preferably "cold" i.e. velocity of the dominant part of the dark matter particles is much less than the speed of light. Though the $\Lambda$CDM model receives an indisputable success on large scales, the validity of the CDM scenario on galactic scales has been questioned in several works. It is found from N-body numerical simulations that CDM halos and sub-halos should have a high density (cuspy) profile at the center \cite{moo94, flo94, deb10}. The CDM model also gives an overabundance of dwarf galaxies in the Milky Way and other similar galaxies/local groups against the observations, which is the so-called missing satellites problem \cite{moo99, kly99, zav09}. There are other issues like the so-called too-big-to-fail problem \cite{boy11, pap15}. However, recent studies claim that except for the core-cusp problem, other discrepancies between observations and CDM-based simulations are removed when baryonic effects are properly taken into consideration in the simulation \cite{saw15}. The warm-cold matter (WDM) has been proposed in the literature as an alternative to CDM \cite{bod01, han02} but the WDM model also shares the core-cusp problem in galactic scale \cite{sch14}. Besides, high redshift Lyman-$\alpha$ forest data disfavors the WDM model \cite{vie13}. There is also a possibility that the core-cusp problem originated due to our poor understanding of galaxy formation or due to improper underlying assumptions in the N-body simulations \cite{wei13, bau18}. Theoretically, weakly-interacting massive particles are the most attractive dark-matter candidates from a particle physics point of view which falls under the cold dark matter category. Considering all the aspects and the overall performance over large scales and galactic scales CDM model still remains the most favored dark matter model. We shall, therefore, derive  $g_{rr}$ from the Einstein field equation considering that the pressure of dark matter is negligibly small i.e. considering essentially the energy-momentum tensor of cold dark matter. 

Considering that the dark matter as a fluid with energy density $\rho(r)$, radial pressure $p_r(r)$, and tangential pressure $p_T(r)$, the Einstein field equations for dark matter halo read (we shall take $c=1$ throughout the manuscript) :

\begin{equation}
\frac{2m^{\prime}(r)}{r^{2}}=8\pi \rho
\end{equation}

\begin{equation}
\frac{2}{r^2} \left[ r \lambda^{\prime}(r) \left(1-\frac{2m(r)}{r}\right) - \frac{m(r)}{r} \right] =8\pi p_r
\end{equation}

\begin{eqnarray}
\left(1-\frac{2m(r)}{r}\right) \left(\lambda^{\prime \prime}(r) + \lambda^{\prime \;2}(r) + \frac{\lambda^{\prime}(r)}{r} \right) \nonumber \\    -\frac{1}{r^3}\left[m^{\prime}(r)-m(r) \right] \left[1+r\lambda^{\prime}(r) \right]  =  8\pi p_T.
\end{eqnarray}

In the galactic halo region $ m(r)>>M_B$ and for cold dark matter $p_r=p_T=0$. Inserting the flat rotation curve led metric coefficient $e^{\lambda}$ i.e. expression given in Eq.(3) in Eqs. (5) and (6) we get to the accuracy of $\beta^{2}_{\varphi}$ 
\begin{equation}
m(r) \simeq  \beta^{2}_{\varphi} r 
\end{equation}
Since the non-trivial vacuum solution is $m_{vac}(r) = M_B$, the complete solution is $m(r) \simeq  \beta^{2}_{\varphi} r + M_B$ (neglecting the higher order terms in $\beta^{2}_{\varphi}$ and cross terms $M_B \beta^{2}_{\varphi}$). The above expression together with Eq. (3) completely specify the halo space-time geometry.

\subsection{Matching with the exterior Schwarzschild space-time}
In general relativity the Schwarzschild metric is the unique static vacuum solution and thus represents the exterior space-time of galaxies with mass parameter equals to total mass $M_T$ content of the galaxy. The solution derived above must match the exterior Schwarzschild metric at the galactic boundary. We consider the junction conditions given by O'brien and Synge \cite{obr52, syn60}  i.e. the metric tensor and all the first order partial derivatives $\frac{\partial g_{\mu\nu}}{\partial x^{\zeta}}$ except possibly $\frac{\partial g_{r\nu}}{\partial r}$ should be continuous at the junction. Note that a solution satisfying the junction conditions of O'brien and Synge always can be transformed to one satisfying the conditions of Lichnerowicz \cite{lic55} and vice versa \cite{isr58, rob72}. 

The matching of metric tensor $g_{tt}$, $g_{ii}$ (i=1,3) and $\frac{\partial g_{tt}}{\partial r}$ at galactic boundary ($r=R_G$, where $R_G$ is the radius of the galaxy) consistently suggest that $C1= 2\beta^2_\varphi (1-ln R_G)$ and

\begin{equation}
M_T = M_B + \beta^2_\varphi R_G . 
\end{equation}

The matching of $\frac{\partial g_{rr}}{\partial r}$ at galactic boundary can be achieved by a coordinate transformation as demonstrated in \cite{rob72} for general class of solutions.   

\subsection{The Tully-Fisher relation}
Tully \& Fisher first demonstrated that an empirical power law relation exists between luminosity and rotation velocity of galaxies \cite{tul77}. However, the optical Tully-Fisher relation exhibits a break; the power law index differs for fainter and brighter galaxies \cite{mcg00}. The rotational velocity of galaxies is found to exhibit a single power law relation with total baryonic disk mass, which is the sum of stellar mass and gas mass of galaxy, instead of luminosity. The baryonic Tully-Fisher relation is given by $ M_B \propto \beta_{\varphi}^{4} $. Our objective is to  replace the rotational velocity by baryonic mass in our final expression for space-time geometry. 

A sample of rotational velocity data for galaxies with large variation in mass are shown in Table 1 which are taken from  \cite{bot85, sch97, san02}. The variation of $\beta_{\varphi}^{4}$ with $M_B$ from the observed data is shown in Fig. (1). Expressing the rotational velocity as $\beta_{\varphi}^{4} = a_{tf}^2 M_B$ where $a_{tf}^2$ is a  proportionality constant, we get  by the least square fitting of the data $a_{tf}^2 =2.43 \times 10^{-24}$ $M_{\odot}^{-1}$. Accordingly, the Eq.(8) reduces to 
\begin{equation}
M_T = M_B + a_{tf} R_G M_B^{1/2}. 
\end{equation}

\begin{table}
\caption{Rotational velocity and baryonic mass data of different galaxies}
\centering
{\scriptsize %
\begin{tabular}{ccccc}
\hline
\hline
\\
Galaxy & $v(kms^{-1})$ & $\delta v(kms^{-1})$ & $M_{stellar}(10^{10}M_{\odot})$  & $M_{gas}(10^{10}M_{\odot})$ \\
\\
\hline
\\
UGC 2885	& 298 & 1	 & 30.8  & 5 \\
NGC 2841	& 287 &	4  & 32.3  & 1.7 \\
NGC 6674	& 242 &	2  & 18    & 3.9\\
NGC 3992	& 241 &	4  & 15.3  & 0.92 \\
NGC 5533	& 240 &	6  & 19    & 3	\\
NGC 7331	& 238 &	2  & 13.3	 & 1.1\\
NGC 3953	& 223 &	3  & 7.9	 & 0.27\\
NGC 801	  & 216 &	1  & 10	   & 2.9\\
NGC 5907	& 215 &	2  & 9.7	 & 1.1\\
NGC 2998	& 212 &	3  & 8.3	 & 3  \\
NGC 5371	& 208 &	2  & 11.5	 & 1  \\
NGC 5033	& 195 &	4  & 8.8	 & 0.93\\
NGC 3521	& 190 &	15 & 6.5	 & 0.63\\
NGC 4157	& 185 &	1  & 4.83	 & 0.79\\
NGC 2903	& 181 &	4  & 5.5	 & 0.31\\
NGC 4217	& 178 &	1  & 4.25	 & 0.25\\
NGC 4013	& 177 &	7  & 4.55	 & 0.29\\
NGC 3893	& 176 &	9  & 4.2	 & 0.56\\
NGC 4088	& 170 &	5  & 3.3	 & 0.79\\
NGC 3877	& 170 &	1  & 3.35	 & 0.14\\
NGC 3726	& 166 &	2  & 2.62	 & 0.62\\
NGC 3949	& 165 &	10 & 1.39	 & 0.33\\
NGC 4100	& 160 &	9  & 4.32	 & 0.3\\
NGC 6946	& 160 &	4  & 2.7	 & 2.7\\
NGC 4051	& 160 &	7  & 3.03	 & 0.26\\
NGC 3198	& 149 &	1  & 2.3	 & 0.63\\
NGC 2683	& 155 &	5  & 3.5	 & 0.05\\
NGC 3917	& 137 &	1  & 1.4	 & 0.18\\
NGC 4085	& 136 &	6  & 1	   & 0.13\\
NGC 2403	& 134 &	1  & 1.1	 & 0.47\\
NGC 3972	& 133 &	3  & 1	   & 0.12\\
UGC 128	  & 130 &	2  & 0.57	 & 0.91\\
F568-V1 	& 124 &	5  & 0.66	 & 0.34\\
NGC 4010	& 123 &	3  & 0.86	 & 0.27\\
NGC 3769	& 118 &	5  & 0.8	 & 0.53\\
NGC 6503	& 115 &	1  & 0.83	 & 0.24\\
NGC 1003	& 112 &	3  & 0.3	 & 0.82\\
F568-3 	  & 111 &	4  & 0.44	 & 0.39\\
NGC 4183	& 111 &	2  & 0.59	 & 0.34\\
F563-V2 	& 111 &	5  & 0.55	 & 0.32\\
F563-1 	  & 111 &	1  & 0.4	 & 0.39\\
UGC 6917	& 111 &	5  & 0.54	 & 0.2\\
UGC 6930	& 110 & -	 & 0.42	 & 0.31\\
M 33	    & 107 & -	 & 0.48	 & 0.13\\
UGC 6983	& 108 &	2  & 0.57	 & 0.29\\
NGC 247	  & 106 &	2  & 0.4	 & 0.13\\
NGC 7793	& 95  &	10 & 0.41	 & 0.1\\
NGC 300	  & 90  &	-  & 0.22	 & 0.13\\
NGC 5585	& 90  &	2  & 0.12	 & 0.25\\
NGC 55	  & 86  &	-  & 0.1	 & 0.13\\
UGC 6667	& 85  &	3  & 0.25	 & 0.08\\
UGC 2259	& 86  &	-  & 0.22	 & 0.05\\
UGC 6446	& 83  &	3  & 0.12	 & 0.3\\
UGC 6818	& 72  &	6  & 0.04	 & 0.1\\
NGC 1560	& 77  &	1  & 0.034 & 0.098\\
IC 2574	  & 68  &	5  & 0.01	 & 0.067\\
DDO 170	  & 64  &	-  & 0.024 & 0.061\\
NGC 3109	& 66  &	3  & 0.005 & 0.068\\
DDO 154	  & 48  &	6  & 0.004 & 0.045\\
DDO 168	  & 54  &	2  & 0.005 & 0.032\\
\hline
\end{tabular}%
}%
\label{Table-1}
\end{table}

\begin{figure}
\begin{center}
\includegraphics[width=0.75\textwidth]{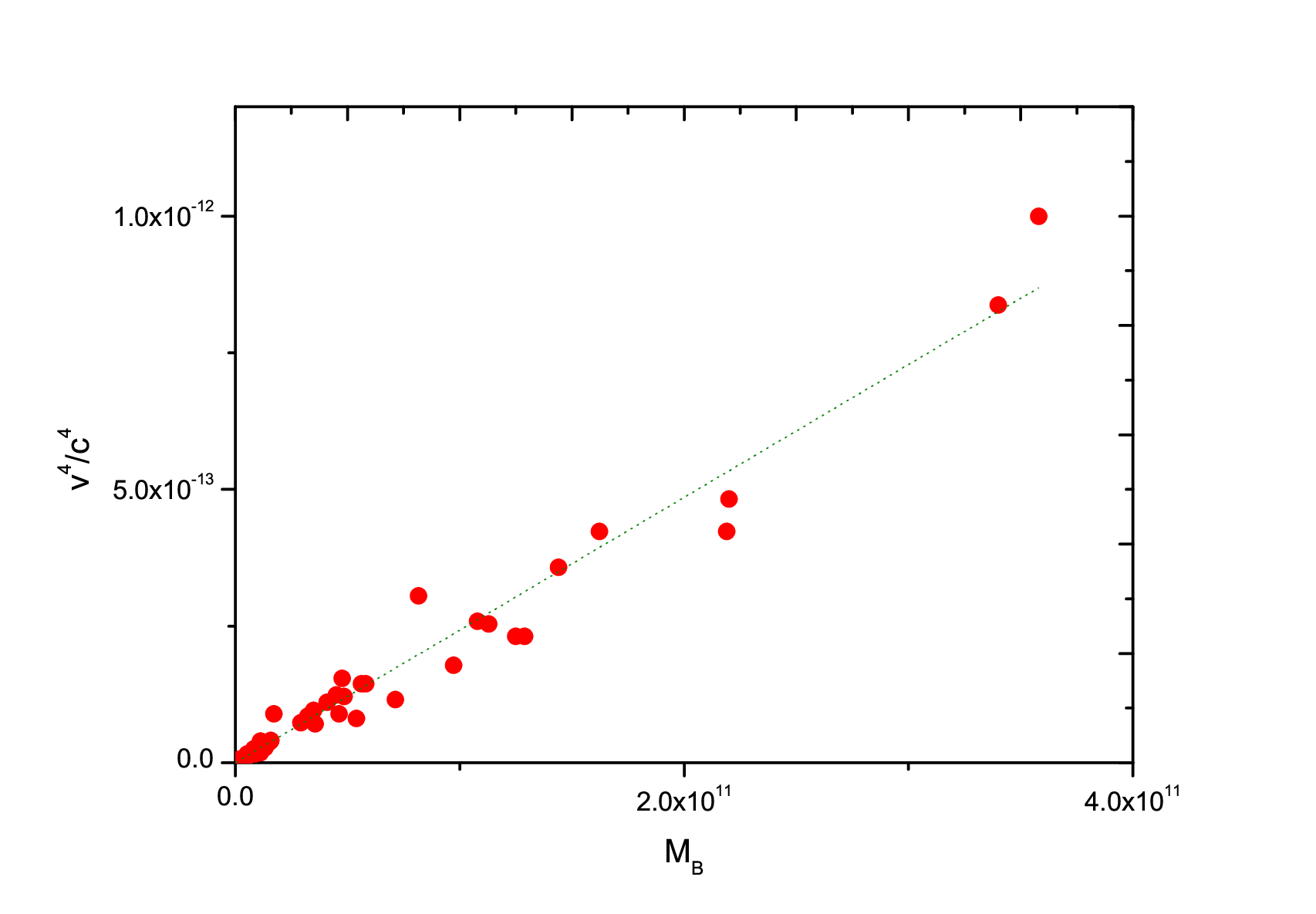}
\end{center}
\caption{The variation of $\beta^{4}_{\varphi}$ with $M_B$ for several galaxies. The dotted line represents the least squared fit of the data.} 
\label{fig2}%
\end{figure}

\subsection{The space-time geometry of galactic halo}
Thus the Eqs. (1), (3), and (7) finally lead the space-time metric of galactic halo 

\begin{eqnarray}
ds^{2} &=&-\left(1-\frac{2M_B}{r} + 2a_{tf} M_B^{1/2} (1+ ln (r/R_G)) \right) dt^{2} \nonumber \\
&& + \frac{dr^2}{1-2a_{tf} M_B^{1/2} -\frac{2M_B}{r}}+ r^2(d\theta^2+sin^2\theta d\phi^2) \;.
\end{eqnarray}
  
The role of Eqs.(8) and (9) is only to fix the constants of Eq.(3). The solution contains a new universal constant $a_{tf}$ which is similar to the universal constant in the MOND scheme \cite{mil83}.  

Instead of flat rotation curve i.e. instead of constant $\beta_{\varphi}$ one may use a few other forms of velocity profiles \cite{per91, gong20}. If we use the universal halo velocity profile as given below \cite{per91,sal07}
\begin{equation}
\beta^{2}_{\varphi} = k \frac{r_a^2}{r_a^2+r^2} 
\end{equation}
where $\frac{\rho_o}{4 \pi}$, $\rho_o$ is the central density and $r_a $ is a constant, the potential f(r) will become

\begin{equation}
f(r) \simeq k ln [(r_a^2+r^2)/R^2] 
\end{equation} 
which reduces to Eq.(3) when $r>>r_a$, with $k=\beta^{2}_{\varphi}$. 

Numerical simulations suggest an approximate universality for the density profile of cold dark matter halos \cite{nav96}. The density profile of dark matter prescribed by Navarro, Frenk and White (NFW) is widely used which is given by \cite{nav96} 

\begin{equation} 
\rho(r) = \frac{\rho_s}{r/r_s (1+r/r_s)^2}
\end{equation}

The mass function corresponds to the NFW density profile will be
\begin{equation} 
m_{NFW}(r) = 4\pi r_{s}^3 \rho_s \left(\frac{r_s}{r + r_s} + Log(r + r_s) \right)
\end{equation}

However, for $m_{NFW}(r)$ (together with $\lambda$) the radial and transverse pressure of dark matter fluid do not vanish as expected for cold dark matter. The assumption of exactly flat rotation curve in deriving metric potential can be the reason for such an inconsistency.  

\section{Stability of the orbits and other general features}

For the metric given in Eq. (1)  circular orbits will exist when $0<r\lambda^\prime <1$ which is indeed the case for the solution given in Eq. (10). The time-like circular geodesics has to be stable for a viable space-time geometry of the galactic halo. The condition for stable circular orbit for the metric given in Eq. (1) is \cite{lak04}
\begin{equation}
3\lambda^\prime + r\lambda^{\prime \prime} >2 r\lambda^{\prime \; 2} 
\end{equation} 
As $\beta <1$, the above condition satisfies for the derived metric. 
  
Inserting the solution of $m(r)$ in Eq. (4), the density of dark matter is readily obtained as 

\begin{equation}
\rho \simeq \frac{1}{4 \pi }   \frac{a_{tf} M_B^{1/2}}{r^2}
\end{equation}
At least in the outer part of galaxies dark matter has a mass density profile closely resembling that of an isothermal sphere. 

The total gravitational energy $E_G$ between two fixed radii, say $r_i$ and $r_o$ in the halo region can be estimated for the metric (Eq.(10)) following \cite{lyn07} which is given by

\begin{equation}
E_{G} = M_{DM}-E_{M}=4\pi\int_{r_{i}}^{r_{o}}\left[1-\left(1-\frac{2m(r)}{r} \right)^{-1/2}\right]\rho r^{2}dr
\end{equation} 

where $M_{DM}$ is the dark matter mass (we have ignored the contribution of luminous matter) which is given by 
\begin{equation}
M_{DM}=4\pi\int_{r_{i}}^{r_{o}}\rho r^{2}dr \simeq a_{tf} M_B^{1/2} \left( r_o-r_i \right)
\end{equation}

The gravitational energy $E_G$ is, therefore, given by

\begin{equation}
E_{G} \simeq - a_{tf}^2 M_B \left( r_o-r_i \right)
\end{equation}

So the first non-vanishing terms is $\mathcal{O}(\beta^4)$.  
%The variation of $E_G$ with radial distance is shown in Figure 1 for Milky Way taking the inner radius as 20 kpc.  

%\begin{figure}
%\begin{center}
%\includegraphics[width=0.5\textwidth]{Graph1.eps}
%\end{center}
%\caption{The variation of $E_{G}$ with ~r~ in Kpc for Milky Way. The lower limit of
%integration i.e. inner radius has taken as 20 kpc.} 
%\label{fig2}%
%\end{figure}

\section{Gravitational lensing due to gravitational field of galactic halo}
%\label{sec:3}
It is often argued that combined observations of galaxy rotation curves and gravitational lensing can provide better insight of the gravitational field of the galactic halo \cite{bha03, fab06}. In \cite{fab06} it was shown that the form of gravitational potential extracted from rotation curve $\lambda_{RC}(r)$ and lensing observations $\lambda_{Lens}(r)$ are not the same in general:

\begin{eqnarray}
\lambda_{RC}(r) = \lambda(r) , \; \nonumber \\
\lambda_{Lens}(r) = \frac{1}{2} \lambda(r) + \frac{1}{2} \int \frac{m(r)}{r^2} dr
\end{eqnarray}

For the halo metric given in (10),  $\lambda_{RC}(r) = \lambda_{Lens}(r) = \lambda(r)$ owing to pressure less fluid. 

In gravitational lensing scenarios when photon trajectories are outside the galaxy, which is the case in most of the observations involving external galaxies/galaxy clusters as a lens, the gravitational deflection will be that due to Schwarzschild geometry with total mass as given by Eq.(9). In such cases, the gravitational lensing phenomenon can provide information about the total mass of the galaxy, check the validity of Eq.(9), and thereby the halo metric. 

When the null geodesics are through a galactic halo the lensing phenomenon may additionally probe the space-time geometry of the halo. When source and observer are at large distance away compare to the distance of the closest approach ($r_o$), for the metric given in Eq.(1) the contribution of the gravitational bending angle over the journey from $r_o$ to infinity can be written as
\begin{equation}
\phi(r_o) - \phi(r_{\infty}) = \int_{r_o}^{\infty} \frac{dr \sqrt{\left(1-\frac{2m(r)}{r}\right)^{-1}}}
{r\sqrt{\left(\frac{r^2}{r_o^2}\left(\frac{e^{2\lambda(r_o)}}{e^{2\lambda(r)}} -1\right)-1\right)}}  %\\ \nonumber
%&&  + \simeq \frac{4 M_B}{R_G}
\end{equation}      

For the halo metric given in Eq. (10), the total bending angle at the leading order in $m/r$ and $\beta_{\varphi}$ will be 
\begin{equation}
\alpha \simeq \frac{4 (M_B + a_{tf} r_o M_B^{1/2})}{r_o}  \; .
\end{equation}      

Therefore, the deflection angle will be enhanced by a factor $4a_{tf} M_B^{1/2} $ over the Schwarzschild value for light trajectory from source to observer. Note that the conventional dark matter model (Newtonian) gives a constant bending angle when the distance of the closest approach of photon trajectories is within the galaxy. Usually, Schwarzschild deflection angle is employed to interpret lensing observations. The Schwarzschild deflection angle ($\frac{4M(r)}{r_o}$) becomes a constant when (dark matter) mass increases linearly with halo radius. Here is a point to be noted: the expression for Schwarzschild deflection angle is evaluated under the assumption that mass parameter $M_T$ is a constant, independent of the radial coordinate. So the application of Schwarzschild deflection angle for dark matter radial dependent mass is not proper as long as the distance of the closest approach is within the galaxy. 

The angular position of the images ($\zeta$) can be obtained from the lens equation in the weak lensing scenario is given by \cite{sch92}\begin{equation}
\zeta = \eta + \frac{d_{ls}}{d_{os}} \alpha
\end{equation}
where $\eta$ denotes the angular source position, $d_{ls}$ and $d_{os}$ are the distances between lens and source and observer and source respectively. The image positions can be obtained from the above equation after inserting the expression for bending angle either from Eq. (22) or the Schwarzschild deflection angle depending on whether the distance of the closest approach is inside or outside the galaxy. 

If the mass of baryonic matter in a galaxy is known independently by some other method such as through photometry, the prediction of halo space-time metric can be tested observationally through lensing observations.  

The expression of image position in weak field Schwarzschild lensing is given by   
\begin{equation}
\zeta_{\pm} = \frac{1}{2} \left(\eta \pm \sqrt{4\alpha_{0} + \eta^{2}} \right)
\end{equation}
where the indices $\pm$ denote the parities of the images, and

\begin{equation}
\alpha_{0} \equiv \sqrt{\frac{d_{ls}}{d_{ol} d_{os}} 4M_T} \;. 
\end{equation}

The next step is to probe the halo space-time geometry through gravitational lensing observations. The gravitational lensing by dark matter haloes (foreground galaxies) induces a coherent distortion to the images of background galaxies (sources). The amplitude of distortion depends on a few factors like the mass (dark matter) distribution in the lens (galactic haloes), angular diameter distances between the lens and the background galaxies, etc. Unless the lens is very massive such as a cluster of galaxies, no reliable estimates of distortion are possible for a single lens-source pair because of the low signal-to-noise (S/N) ratio. Instead, by averaging over all the lenses in a sample, a high (S/N) can be achieved in the estimate of mean shear as a function of the angular separation between the lens and the source, and subsequently, the ensemble-averaged distribution of mass around the foreground galaxies can be computed.

\subsection{Mass-luminosity correlation}
The luminous part of galaxies are supposed to reside in dark matter haloes. The mass density of dark matter in singular isothermal sphere profile as a function of radial distance r is considered as $\rho_{sis} = \frac{\sigma^2}{2 \pi G r^2}$, where $\sigma$ is the velocity dispersion of a test particle. Consequently, the effective dark matter mass in a galaxy within the radius r can be written as
  
\begin{equation}
M_{eff}(<r) = 2\sigma^2 r 
\end{equation}

The velocity dispersion of singular isothermal sphere density profile of galactic halo goes with luminosity (L) of galaxies as $\sigma \propto L^a$ where the exponent $a$ is 0.24 and 0.23 respectively for red  and blue lenses \cite{bri13} as found from the galaxy-galaxy lensing observations by the Canada–France–Hawaii Telescope Legacy Survey Wide (CFHTLS-wide). 

The expression of gravitational bending angle as given by Eq.(22) suggests that the accumulated mass $(M_h)$ within the radius r of a galactic halo is given by 

\begin{equation}
M_h(<r) \sim a_{tf} r M_B^{1/2} 
\end{equation}

Comparing with the Eq.(26) we find that the effective dispersion velocity in our cold dark matter scenario is

\begin{equation}
\sigma_{eff}(<r) =  (\frac{a_{tf}^2 M_B}{4})^{1/4}  
\end{equation}
 
Note that $\sigma_{eff}$ is not the actual baryonic velocity dispersion, but a proxy. In general luminosity of a galaxy goes proportionally with baryonic mass of the galaxy. Hence, for the present model $\sigma_{eff}(<r) \propto L^{0.25}$ which is in accordance with the CFHTLS-wide observations.

\subsection{Lensing observations due to Abell 370 and Abell 2390}

Since the presence of dark matter is clearly revealed in galaxy clusters, here we have considered the case of gravitational lensing by galaxy cluster Abell 370 and Abell 2390. In Abell 370 the \lq giant luminous arcs\rq were first observed \cite{lyn86}, \cite{sou87}. Our objective is to examine whether the non-baryonic part ($\beta^2_\varphi R_G$) of the total mass as given by Eq.(8) can describe the estimated total mass $M_T$ of the stated two clusters from the lensing observations.  

A light ray suffers gravitational deflection while passing the lens plane at a point $r(x,y)$ from the center of the lens by an angle 

\begin{equation}
\boldsymbol{\alpha(r)} = 4 \int\int dx^{\prime} dy^{\prime} \Sigma(r^{\prime}) \frac{\mathbf{r-r^{\prime}}}{|r-r^{\prime}|^2}
\end{equation}

where $\Sigma(r^{\prime})$ is the surface mass density in the lens plane. If the mass distribution of the lens is spherically symmetric and if the source is located behind the lens along the optic axis then one obtains the radius of Einstein ring from the lens equation 

\begin{equation}
\theta_{E}^2 =  \frac{M_T(d_{ol}\theta_{E})}{\pi \Sigma_c D_{ol}^2}
\end{equation}
 
where $\Sigma_c \equiv  \frac{1}{4\pi} \frac{D_{os}}{D_{ol} D_{ls}}$ is the critical density, $D_{ol}$, and $D_{os}$ are the angular distance of the lens and the source from the observer respectively and $D_{ls}$ is the angular distance of the source from the lens. For a point mass lens or spherically symmetric mass distribution $M_T=M_T(d_{ol}\theta_{E})$. Galaxy clusters have complex matter distributions in general and cannot be considered to be either point masses or spherically symmetric mass distribution. However, a spherically symmetric lens model can be employed as a first approximation to extract the same order of magnitude results as the more realistic case analyzing the large arcs that are observed in clusters \cite{ber90, pel91}. Thus, the total interior mass at the zeroth order accuracy may be read as 

\begin{equation}
M_{T}(d_{ol}\theta_{E})  =  \theta_{E}^2 \frac{d_{ol} d_{os}} {4 d_{ls}}
\end{equation} 

The ratio of mass-to-light ($M_T/L$) within A0 in Abell 370 from photometric measurement is found around 100 \cite{sou87} or even 190 \cite{gro89}. Since for luminous matter $M_T/L$ is just a few $M_{\odot}/L_{\odot}$, the luminous mass of the cluster is at least an order smaller than the total mass of the cluster. In \cite{ber90} the dark matter mass in the source within A0 was estimated as about $ 1 \times 10^{14}$ $M_{\odot}$. 

To describe the above observations we have considered the luminous arc, A0, which has a radius of curvature of about $25^{\prime \prime}$ \cite{gro89} and treat the arc as an Einstein ring \cite{gro89, pac87}. The  observed redshift ($z_{s}$) of A0 is $0.724$ which gives the distance of the background galaxy and the lens distance is obtained from the redshift $0.374$ of Abell 370. A concordance cosmological model of ($\Omega_m, \Omega_{\Lambda}, \Omega_k$) = $(0.286; 0.714; 0)$ with $H_o=69.6$ is applied for distance estimation from redshifts of lens and source. Subsequently, we have checked whether the non-baryonic part of ($\beta^2_\varphi R_G$) of the total mass as given in Eq.(8) can describe the bulk of the estimated total mass $M_T$ of the system from the lensing observation. Accordingly, we evaluate the value of $\beta_{\varphi} R_E$ for the purpose using the observed $\beta_{\varphi}$ from the velocity dispersion data. We use $\beta_{\varphi}$ as velocity of dispersion for Abell 370 which is $1700 \pm 180$ $km \; s^{-1}$ \cite{hen87, gro89}. The findings are given in the table 2. For Abell 370 our estimated dark matter mass is within $15\%$ of that found in \cite{ber90}. The present model gives the ratio of dark matter mass to total mass of Abell 370 as $0.74$ which is consistent with what was obtained ($\sim 0.7$)in \cite{gro89}.    

We have also considered another cluster Abell 2390 ($z_l=0.231$) in which the photometry and spectroscopy observations reveal a very linear arc having the radius of curvature of about $37^{\prime \prime}$ \cite{pel91}. The observed redshift of the arc is 0.913. The velocity of dispersion was reported in the range $1100$ - $1250 $ $km \; s^{-1}$ \cite{pel91, car96} and we have used the average value of $1175$ $km \; s^{-1}$. The observed mass-to-light ratio suggests the dark matter dominance in the cluster \cite{squ96}. We have estimated the total mass and the dark matter mass of the cluster following Eqs.(31) and (8) respectively which are also shown in the Table 2. It is found that the present model correctly describes the dominance of dark matter contribution in both the galaxy clusters.

\begin{table*}[t] 
%[htbp]
\large
\centerline{\large Table 2}
\centerline{\large Estimated total and dark matter mass of Abell 370 and Abell 2390}
\begin{center}
\begin{tabular}{cccccccccccc}
\hline
\hline
\noalign{\vskip 2mm}
${\rm Object}$ & ${\rm z_{l}}$ & $z_{s}$ & $r_E $ & $M_T/M_{\odot} \times 10^{-14}$  & $M_{DM}/M_{\odot} \times 10^{-14}$    \\
$ $ & $ $ & $ $ & in arcsecs  & from lensing data  & from Eq.(8)   \\
$$  & $ $ & $ $ &             &                    &               \\
\hline
\hline
\noalign{\vskip 2mm}
$Abel 370  \;$ &  $0.374$ & $0.724$  & $25$  & $ 1.15$ & $0.85 $ \\
\hline
$Abel 2390  \;$ &  $0.231$ & $0.913$  & $37$  & $ 0.68$ & $0.63 $  \\
\hline
\hline
\end{tabular}
\end{center}
\end{table*}

\section{Discussion and Conclusion }
%\label{sec:3}
In the present work, the form of the gravitational potential of galactic halo led by the flat rotation curve feature is derived. The mass function is obtained considering the presence of cold dark matter in the galaxy. Note that the mass function will alter from what derived here if any other form of dark matter such as perfect fluid, or scalar field-inspired dark matter state is considered. However, gravitational potential derived from lensing observations for a different choice of dark matter state instead of cold dark matter, in general, does not consistently match with that obtained from the flat rotation curve feature \cite{fab06}.  

There exist several mass models such as the singular isothermal sphere model, singular isothermal ellipsoid model, pseudo-Jaffe lens model and so on which are developed to explain the galactic rotation curves together with the lensing observations \cite{kor94, knu00}. The basic approach of all such models is to formulate the mass distribution of a galaxy in an empirical way so that the rotation curves of spiral galaxies can be described. Subsequently, the surface mass density in the lens plane is calculated, and using that the gravitational lensing properties are studied (for instance employing Eq.(29)). However, these mass models do not give space-time metrics, unlike the present case. In the present work we, instead, obtained the space-time metric of the galactic halo as a solution of cold dark matter equation of state and thereby consistent with the standard concordance cosmological model. In contrast, the mass models are not necessarily consistent with the cold dark matter paradigm. The major merit of the present model is that it provides an opportunity to test the cold dark matter scenario through other local gravitational phenomena like gravitational time delay, perihelion precession, etc. since we have a space-time metric of the galactic halo.              

The radial extension of a galaxy (i.e. of dark matter) is so far not known which is one of the unanswered questions of modern astrophysics. Some alternative theories to GR, particularly conformal gravity can explain the galactic rotation curve without invoking any dark matter component \cite{man11}. However, the radial extent of galaxies in alternative theories \cite{nan12} is not the same as the GR prediction and hence this feature is also a testable observable to differentiate the two models. In the present model, the total mass of dark matter is found proportional to the radius of the galaxy. So if total mass is obtained say from lensing measurements and if the baryonic mass is estimated from say photometric study, one can readily estimate the radial extension of a galaxy from Eq.(9) under the present framework and hence is an important testable parameter for the model.  

The form of gravitational potential implies that at small r the 1/r term will dominate whereas at large r the logarithmic term will command. At the intermediate region there has to be a crossover radius $r_{c}$ where $\beta^2_\varphi(r_{c}) = \frac{M_B}{r_{c}(1+ln r_c/R_G)}$ beyond which the rotational velocity does not change much. The Tully-Fisher relation demands that for each galaxy the crossover radius is roughly proportional to $M_{B}^{1/2}$ which is an interesting corollary of the dark matter hypothesis. 

Gravitational lensing is an important tool to discriminate dark matter and alternative dark matter models. Here we have derived the gravitational lensing formulation for the halo metric and the predictions of the model have been compared with the galaxy-galaxy weak lensing observations and the lensing observation of Abell 370 and Abell 2390 clusters. The present model correctly describes the galaxy-galaxy weak lensing observational feature relating velocity dispersion and galactic luminosity (L) as $\sigma \propto L^{1/4}$. It is found that the derived halo space-time can successfully explain both the aforementioned lensing observations. 

\section*{Acknowledgments}

The authors thank two anonymous reviewers for useful comments that help us to improve the manuscript. The work of AB is supported by Science and Engineering Research Board (SERB), DST through grant number CRG/2019/004944.

\appendix

\section{Appendices}
For a static spherically symmetric metric given by
\begin{equation}
ds^{2}=-w(r) dt^{2}+ q(r) dr^2+ r^{2}(d\theta^2+sin^2\theta d\phi^2),
\end{equation}
where $w(r)$ and $q(r)$ are function of r only, the Einstein's field equations with a fluid ($ G_{\mu\nu} = 8 \pi T_{\mu\nu}$) reduce to \cite{wal84}

\begin{eqnarray}
G_{oo} = (r q^2)^{-1} q^{'} +r^{-2} (1-q^{-1}) = 8\pi T_{oo} = 8\pi \rho
\end{eqnarray}

\begin{eqnarray}
G_{rr} = (r wq)^{-1} w^{'} - r^{-2} (1-q^{-1}) = 8\pi T_{rr} = 8\pi P_{r}
\end{eqnarray}

\begin{eqnarray}
G_{\theta \theta} = \frac{1}{2} (w q)^{-1/2} \frac{d}{dr} [(w q)^{-1/2} w^{'}] + \frac{1}{2} (r wq)^{-1} w^{'} -\frac{1}{2} (r w^2)^{-1} w^{'} &=& 8\pi T_{\theta \theta} \\ \nonumber 
&& = 8\pi P_{T}
\end{eqnarray}

The choice $w(r)=e^{2\lambda(r)}$ and $q(r) =\frac{1}{1-\frac{2m(r)}{r}}$ in the above equations lead to Eqs(4-6).
%\section{References}

\end{document}